\documentclass[12pt,preprint]{aastex}

\newcommand{\etal}{{\it et al.}}

\shorttitle{AN EXTREMELY RED GALAXY AT z=0.65}
\shortauthors{Afonso, Mobasher, Chan, \& Cram}

\begin{document}

\title{Discovery of an extremely red galaxy at z=0.65 with dusty 
star formation and nuclear activity \footnote{Based on observations 
with the {\it Infrared Space Observatory}, Australia Telescope Compact 
Array, Anglo Australian Telescope, Cerro Tololo Interamerican Observatory 
and the European Southern Observatory (prg: 266.A-5633)}}

\author{J. Afonso}
\affil{Blackett Laboratory, Imperial College, 
Prince Consort Rd, London SW7 2BW, UK}
\email{j.afonso@ic.ac.uk}

\author{B. Mobasher\altaffilmark{2}}
\affil{Space Telescope Science Institute, 
3700 San Martin Drive, Baltimore MD 21218, USA}
\email{b.mobasher@stsci.edu}

\and

\author{B. Chan and L. Cram}
\affil{School of Physics, University of Sydney, Sydney NSW 2006, Australia}
\email{bchan@physics.usyd.edu.au, l.cram@physics.usyd.edu.au}

\altaffiltext{2}{Also affiliated to the Space Sciences 
Department of the European Space Agency}

\begin{abstract}
In the course of the follow-up multiwavelength study of a deep radio
survey we have discovered that the milli-Jansky radio source PDFJ011423 is 
a low-redshift ($z = 0.65$) extremely red galaxy (ERG) with $K=15.3$, 
$R-K = 5.8 $ and $J-K = 3.1$. Optical, infrared and radio photometry, 
together with optical and near-infrared spectroscopy, reveal a heavily 
obscured galaxy ($A_V$=5--6, from the observed Balmer decrement) 
undergoing 
vigorous star formation and presenting an active galactic nucleus (AGN). 
PDFJ011423 is a representative member of the dusty ERG population, 
providing a local counterpart for studying more distant ERGs. 

\end{abstract}


\section{INTRODUCTION}

Deep multi-wavelength near-infrared and optical surveys over the last 
decade have led to the discovery of a population of galaxies with very 
red optical-infrared colors \citep[``extremely red galaxies'', hereafter 
ERGs; e.g.][]{Els88,Mcc92,Hu94,Tho99,Dad00}. ERGs, defined as 
galaxies with $R-K > 5$, appear to fall into two sub-classes: (1) high 
redshift ($z \gtrsim\ 1$) evolved elliptical galaxies with intrinsically 
red spectral energy distributions (SEDs) 
\citep[e.g.][]{Spi97,Sti99,Soi99} and (2) highly obscured dusty galaxies 
undergoing starburst activity \citep[e.g.][]{Cim98,Dey99,Sma99}. 
The predominance 
in the ERG population of the latter would favor hierarchical galaxy 
formation scenarios  \citep{Whi91}, while the former would support early 
formation of galaxies through monolithic collapse \citep{Lar75}. 
Therefore, the study of ERGs can help constraining existing models for 
formation and evolution of galaxies. 

The observed very red colors of those ERGs which are classified as 
elliptical galaxies are attributed predominantly to the large K-correction 
arising from their high redshift. On the other hand, star-forming galaxies 
will owe their redness predominantly to dust extinction. This 
view implies that extremely red ellipticals will occur only at high 
redshift ($z \gtrsim\ 1$) while extremely red galaxies at $z \lesssim 1$ 
will exhibit unusually heavy obscuration of either stellar or AGN 
emission, or both. This is expected to be the reason for the 
small space density found for ERGs in the local Universe.

This letter reports the discovery of an ERG, PDFJ011423, at $z = 0.65$. 
Multi-waveband photometric and spectroscopic observations reveal 
evidence of both star formation and AGN activity in this galaxy, which 
can be regarded as a ``local'' template for the study of this class of 
ERG at higher redshifts. 
Throughout this paper we adopt 
$H_0 = 65~h_{65}$ km\,s$^{-1}$\,Mpc$^{-1}$ and $q_0=0.5$.

\section{OBSERVATIONS}

PDFJ011423 ($\alpha = 01^{h}14^{m}23^{s}$, $\delta =-45^{\circ}34'30''$, 
J2000) was first noted as a faint 1.4\,GHz radio source 
($S_{1.4\,\rm{GHz}}=1.67$\,mJy) in the Phoenix Deep Survey
\citep{Hop98,Hop99}. Aperture photometry of the optical counterpart of 
the source gives $R$=21.1 and $V$=22.7 \citep{Geo99}, locating 
PDFJ011423 close to the magnitude limit of the follow-up Two Degree Field 
multi-fibre spectroscopic survey \citep{Geo99}. Nevertheless, a 
strong emission line, identified as [OII]3727  at a redshift 
$z=0.65$, was detected.

In the light of the known strong correlation between the far-infrared
(FIR) and 1.4\,GHz luminosities for star-forming galaxies, these clues 
were sufficient to schedule PDFJ011423 for observation by the {\it 
Infrared Space Observatory} ({\it ISO}), using ISOCAM 
(7 and 15\,$\mu$m) 
and ISOPHOT (90\,$\mu$m). The {\it ISO} observations (J. Afonso \etal,
in preparation) revealed the presence of a relatively bright source at 7 
and 15\,$\mu$m, with fluxes of 4.1 and 7.6\,mJy, respectively. The object 
was also detected by {\it ISO} at 90\,$\mu$m (4$\sigma$ level), with a 
flux of 260\,mJy.  Subsequent near-IR photometry, using the 
New Technology Telescope (NTT) and the 
CTIO 1.5m telescopes, showed this galaxy to have $K$=15.3 and $J$=18.4. 
PDFJ011423 is thus classified as an ERG with $R-K=5.8$ and $J-K=3.1$. 

The closeness and the relatively bright optical luminosity of PDFJ011423 
allows detailed spectroscopy of this object at both optical and near-IR 
wavelengths. The optical spectroscopic observations were obtained at the
Very Large Telescope (VLT) using the FORS1 instrument in longslit mode. 
Three dithered observations of 15 minutes each were made using a 1.6$''$ 
wide slit and a low resolution grism (150I+17), covering the wavelength 
range $\lambda \lambda$0.6--1.1 $\mu$m. Bias subtraction, flatfielding and 
the combination of the dithered observations were performed before 
extracting the final spectrum using the software package IRAF. The 
wavelength scale was defined by fitting a third order polynomial to the 
lines of a He-Ar 
calibration spectrum. The spectrum of the standard star GD50 was taken
for the flux calibration, which agrees with the optical photometry of 
PDFJ011423 within 15\%. Near-infrared spectroscopy was obtained using SOFI on 
the ESO New Technology Telescope. Twelve observations of 270 seconds 
each, dithered along a  1.0$''$ slit, were obtained. The low resolution 
grism covering the range $\lambda \lambda$0.95--1.64 $\mu$m was used. 
The dithered observations were flatfielded and combined before 
the spectrum was extracted. For the wavelength calibration a 
Xe lamp spectrum was used. 
The photometric calibration was performed using the $J$-band magnitude 
of the galaxy. The final calibrated spectra are presented in 
Figure~\ref{fig:spectra} with the respective line measurements, 
based on Gaussian fits to the emission lines, given in Table~\ref{tab:lines}. 

\section{NATURE OF PDFJ011423}

\begin{figure}
\epsscale{0.95}
\plotone{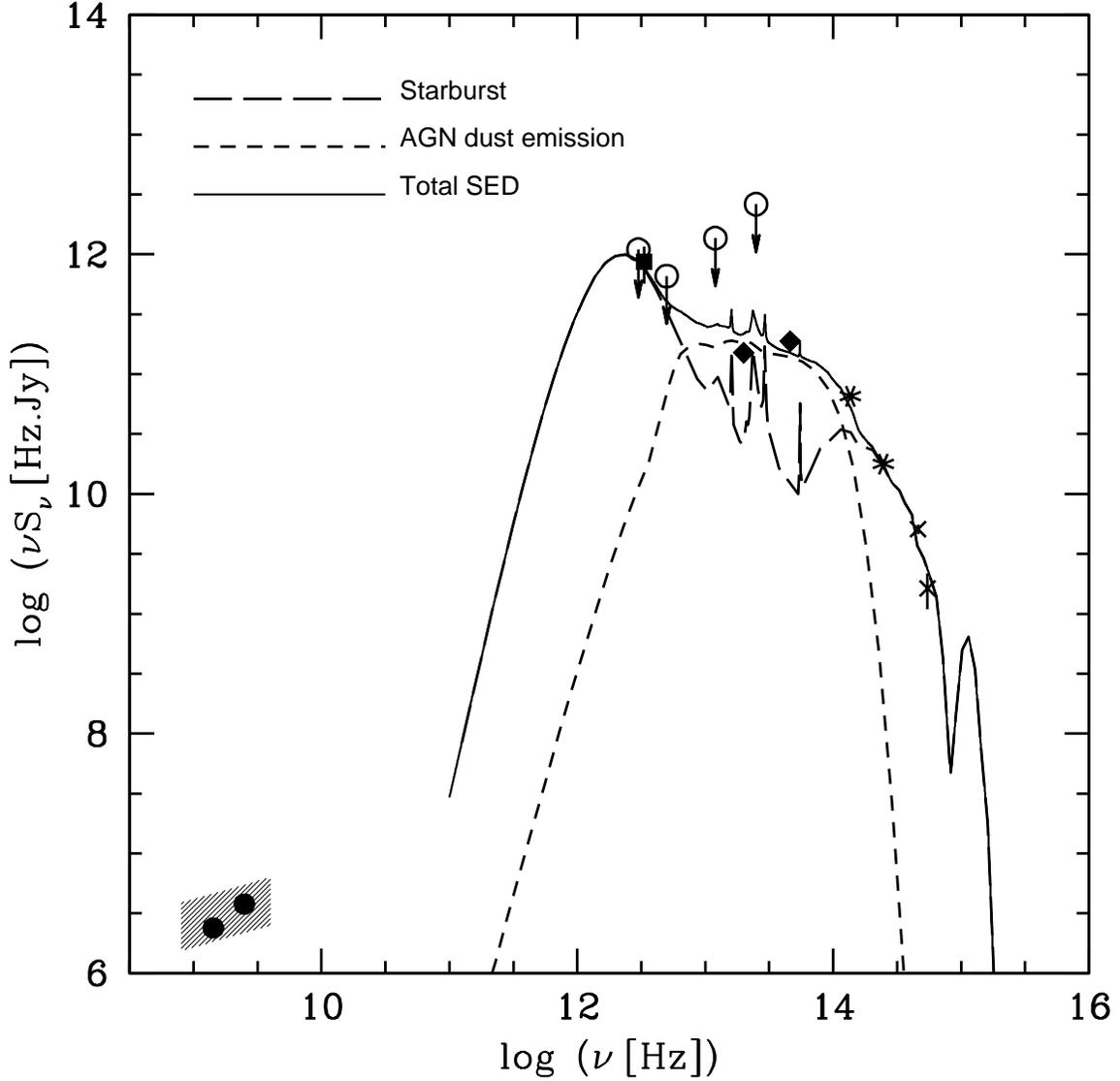}
\caption{Observed SED of PDFJ011423
compared with the fitted starburst (long-dashed line) and AGN dust
emission (short-dashed line) models. Crosses and stars denote, 
respectively, the optical ($V$ and $R$-bands) and near-infrared 
($J$ and $K$-bands) measurements. Detections at mid and far-infrared 
are from ISO (filled diamonds and square) while upper limits 
(3$\sigma$, open circles) are shown for co-added observations from 
the {\it Infrared Astronomical Satellite} (IRAS) at  
$\lambda_{obs} =$ 100, 60, 25, and 12\,$\mu$m, 
estimated using the SCANPI procedure at the Infrared
Processing and Analysis Center. Filled circles are the radio detections 
at 1.4 and 2.4\,GHz. The shaded region represents the
prediction of the radio flux based on FIR/radio correlation described
in the text.} 
\label{fig:sed}
\end{figure}

We approach the interpretation of the observed spectral energy distribution
(SED) of PDFJ011423 
(Figure~\ref{fig:sed}) by attempting first to model its continuum 
emission, and then checking for consistency with the observed spectral 
line emission. The SED of a normal {\it elliptical} galaxy can only 
reproduce the observed very red colors of PDFJ011423 at $z\gtrsim\
1$. This contradiction with the known redshift, combined with the 
presence of strong emission lines, eliminates PDFJ011423 as an 
elliptical galaxy. 

The high luminosity of the galaxy at rest-frame 60\,$\mu$m suggests the 
presence of a dusty starburst, while not excluding the possibility that 
some of the near-IR and mid-IR luminosity arises from nuclear activity. 
Accordingly, we have matched the observed fluxes by combining the model 
SEDs of a starburst occurring in evolving, 
dusty giant molecular clouds (GMCs) \citep{Efs00} and of a 
dust-enshrouded AGN \citep{Row95}. Although it is 
possible that the two energising phenomena are physically linked 
\citep[e.g.][]{Row95}, here we simply superimpose the SEDs predicted by 
the two models. 

The model fit, presented in Figure~\ref{fig:sed}, results from a
$\chi^2$ minimization to the observed SED, varying within the domain of 
the parameters of the original starburst and AGN models. The more 
powerful component, responsible for 76\% of the infrared luminosity
($L_{1-1000\,\mu m}$), is a model of a dusty starburst of age 57\,Myr, 
with optical opacities of the GMCs in the range 
$\tau_{\rm v} =3-200$. The properties of this component are determined 
primarily by the constraints imposed by the FIR and 
optical emission. 

The starburst component alone is, however, unable to fit the mid-IR 
data. Therefore, an extra contribution in the form of 
emission from hot dust (T $\sim 160 - 1600$\,K) heated by a putative AGN 
is required. This component represents 24\% of the total infrared 
luminosity. Acceptable fits could also be obtained using other AGN 
models, for example by allowing for emission from a dusty torus 
surrounding the central (black hole) source \citep{Efs95}. Both classes 
of models lead to similar descriptions of the object in the context 
of this paper.

The best fitted model SED in Figure~\ref{fig:sed} predicts an infrared 
luminosity of 
L$_{1-1000\,\mu\rm{m}} = 7.1 \times 10^{12}~h_{65}^{-2}$\,L$_{\sun}$, 
which classifies PDFJ011423 as an ultra-luminous infrared galaxy (ULIRG). 
The possibility that the high luminosity is due to gravitational lensing 
is unlikely, given the relatively low redshift of the object. 

The existence of only one data point in the FIR/sub-mm part of the SED 
makes an estimate of the dust mass for PDFJ011423 highly uncertain. 
However, the fitted SED in this wavelength range is compatible with an 
optically thin thermal dust emission spectrum 
(emissivity index of 1.5) with a temperature of 
T$_{\rm dust}= 33 \pm 3$\,K, which corresponds to a dust mass of M$_{dust} 
\sim 8-30 \times\ 10^{8}~h_{65}^{-2}$\,M$_{\sun}$ \citep{Hil83}. 
Consistent with the 
extreme reddening of the galaxy, this value is an order of magnitude 
higher than the dust mass of $10^7-10^8~h_{65}^{-2}$\,M$_{\sun}$ found 
for local ULIRGs \citep{San96} and comparable to the higher values 
found in the sample of PG quasars of \citet{Haa00}, for example.

The two models used to fit the observed SED do not predict the radio 
continuum luminosity or spectral index. To interpret the radio data, we 
appeal to the remarkably tight empirical correlation between the FIR and 
radio continuum luminosity observed in star-forming galaxies 
\citep{Hel85,Con92}. The favored explanation of the correlation holds 
that the same massive stars warm the FIR-emitting 
dust and energise, through supernova explosions, the relativistic 
electrons responsible for the radio continuum. Given the FIR luminosity 
of PDFJ011423, this correlation predicts a 1.4\,GHz flux in the range 
1.3 -- 3.3\,mJy 
compared with the measured value of 1.67 mJy. This agreement 
supports the possibility that most of the radio emission comes from 
star formation processes. It should however be noted that a similar 
radio/FIR correlation holds for samples of radio-quiet quasars, perhaps 
as a consequence of links between star formation and black hole feeding 
rates. The relatively flat radio spectral index observed here, 
$\alpha_{1.4}^{2.4} = 0.16$ (S$_{\nu} \propto \nu^{-\alpha})$, while 
being frequently linked to quasar emission \citep[e.g.][]{Web95}, 
has also been observed in 
compact starburst nuclei of ULIRGs \citep{Sop91,Cra96}. The available 
radio images do not have sufficient spatial resolution to reveal any 
structure in PDFJ011423 and cannot aid to clarify the origin of the  
radio emission. 

\begin{figure}
\epsscale{1.0}
\plotone{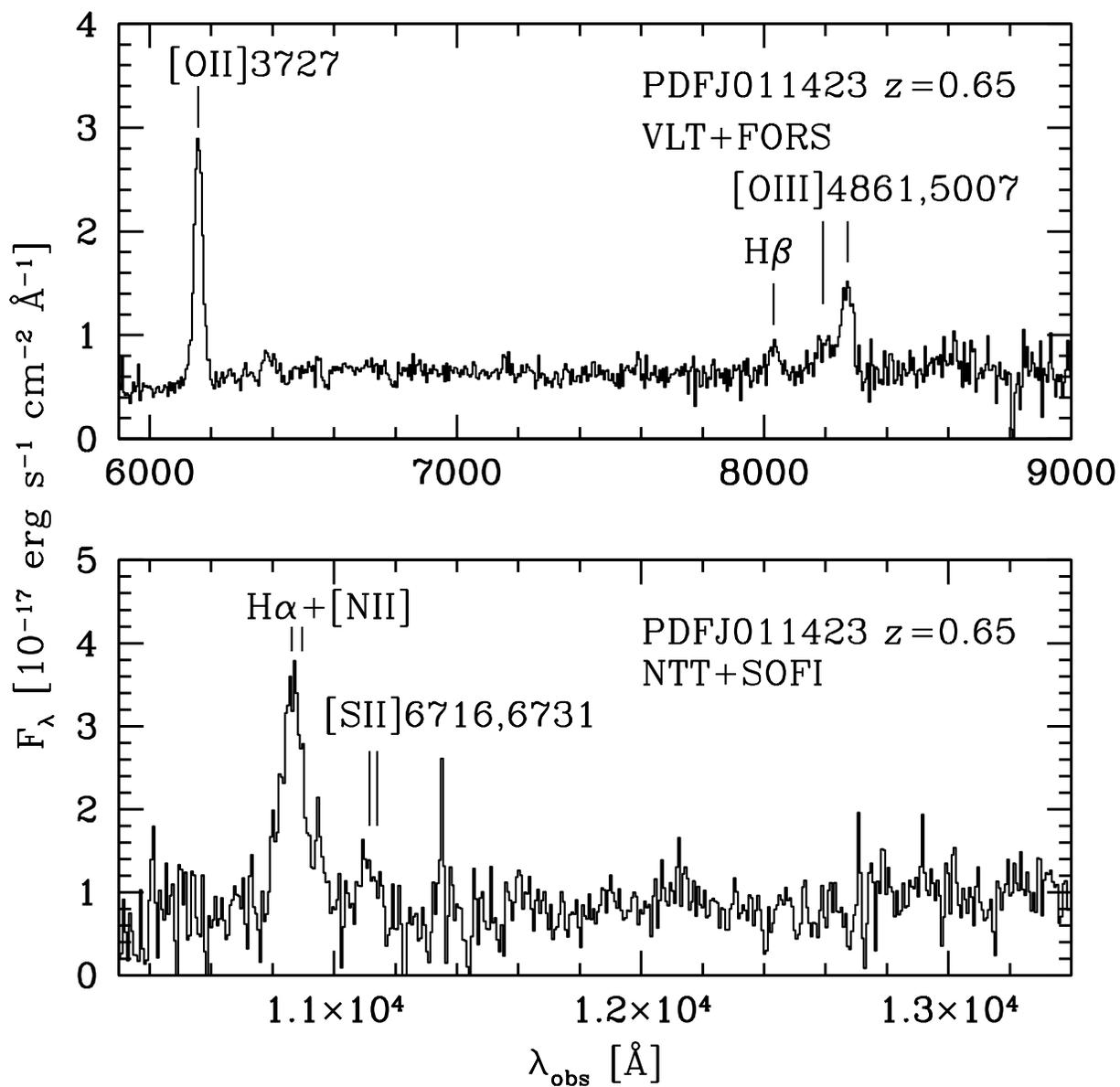}
\caption{Optical and near-infrared spectra of
PDFJ011423 at $z=0.65$. The optical spectrum was obtained at the Very
Large Telescope (VLT) using FORS1 while the near-infrared spectrum 
was obtained at the New Technology Telescope (NTT) using SOFI.}
\label{fig:spectra} 
\end{figure}

We now consider the spectral properties of PDFJ011423, using optical and 
near-IR line diagnostics from Figure~\ref{fig:spectra}.  Since the 
H$\alpha$+[NII] blend is not resolved by the NIR spectrum, 
we assume the H$\alpha$/[NII] ratio to have a value between 1 
\citep[typical of Seyfert galaxies, e.g.][]{Vei87} and 2 
\citep[corresponding to star-forming galaxies,][]{Ken92}
in order to separate the two 
contributions. With the de-blended H$\alpha$ line intensity, the 
observed Balmer line ratio becomes H$\alpha$/H$\beta = 15-20$. Assuming 
Case B Balmer recombination (adopting the intrinsic value for 
H$\alpha$/H$\beta$ as 2.85 for star-forming galaxies \citep{Bro71} and 
3.1 for narrow line AGNs \citep{Vei87}) and a standard Galactic 
extinction curve \citep[][using $R_V=3.1$]{Car89}, we estimate an 
optical dust extinction of $A_V = 5-6$ mag. 
This large extinction is likely to be the source of 
the extreme red color in PDFJ011423. The absence of the 4000\ \AA\ break 
in the optical spectrum, indicating an optical continuum dominated by 
very young stars, is consistent with the inference that this is a
vigorously star-forming galaxy. The measured optical line widths are not 
significantly larger than the instrumental resolution, placing an upper 
limit of approximately 500\,km s$^{-1}$ on their intrinsic value. 
In the NIR spectrum, the H$\alpha$+[NII] blend presents a width value of 
(FWHM) $\sim 2000$\,km s$^{-1}$ (after correction for 
instrumental resolution). Given the low resolution of the spectrum it 
is not clear if this shows an intrinsic broad line component 
for H$\alpha$ or is due to significant contribution from [NII].

The diagnostic line ratios \citep{Rol97,Vei87} for PDFJ011423, after 
correction for dust obscuration, are consistent with the two 
component model. While ratios involving the [OII]3727 line 
([OII]3727/H$\beta$ = 8.0), indicate excitation by an AGN, the other 
line ratios ([OIII]5007/H$\beta$=2.9, [SII]/H$\alpha$=0.2--0.3) and the 
non-detection of [OI]6300 are indicative of a starburst. This kind of 
ambiguous spectral line classification has been linked previously to 
galaxies composite in nature, i.e., hosting both a starburst and an AGN 
\citep{Hil99}.

The presence of an AGN in addition to the starburst component 
complicates the estimation of the star formation rate (SFR). The most 
reliable measure is from its 60\,$\mu$m luminosity ($2.4 \times\ 
10^{26}~h_{65}^{-2}$\,W\,Hz$^{-1}$) where the model implies that the 
dominant contribution is due to the starburst. Using the appropriate 
calibration \citep{Cra98} for an initial mass function (IMF) of the form 
$\Psi (\rm{M}) \sim \rm{M}^{-2.5}$ with stellar masses between 
0.1 and 100 
M$_{\sun}$, we estimate a SFR for massive stars of SFR$_{60\,\mu 
\rm{m}}$ ($\rm{M} \geq 5\,\rm{M}_{\sun}$)$=470~h_{65}^{-2}$ 
M$_{\sun}$ yr$^{-1}$.  
Alternatively, from the observed radio power (P$_{1.4\,\rm{GHz}} = 1.5 
\times\ 10^{24}~h_{65}^{-2}$\,W\,Hz$^{-1}$) we deduce 
SFR$_{1.4\,\rm{GHz}}$ ($\rm{M} \geq 5\,\rm{M}_{\sun}$)$=377~h_{65}^{-2}$ 
M$_{\sun}$ yr$^{-1}$, consistent with the value estimated from the 60 
$\mu$m luminosity. On the other hand, the extinction-corrected 
H$\alpha$ luminosity 
(L$_{\rm{H}\alpha} = 0.9-2.6\times 10^{37}~h_{65}^{-2}$ W) 
implies 
SFR$_{\rm{H}\alpha}$($\rm{M} \geq 5\,\rm{M}_{\sun}$)$=620-1730~h_{65}^{-2}$ 
M$_{\sun}$ yr$^{-1}$, the large range being due to the unresolved 
contribution of [NII] to the observed H$\alpha$+[NII] blend. This 
excess over the SFR estimated from FIR and radio luminosities may
 be due to the AGN contribution. 

Three other dusty ERGs have been spectroscopically confirmed and are 
relatively well studied: HR10, displaying strong starburst activity at 
$z=1.44$ \citep{Dey99}; ISOJ1324-2016, a $z=1.50$ galaxy which hosts a 
dusty quasar \citep{Pie01} and EROJ164023, which at $z=1.05$ shows 
star-forming activity with a possible weak AGN \citep{Smi01}. 
A brief comparison of some
of the observed quantities in PDFJ011423 and these three other dusty ERGs 
is listed in Table 2. Although the non detection of star formation in 
ISOJ1324-2016 can be due to the lack of FIR/sub-mm observations, it seems 
clear that different degrees of AGN and starburst activity do exist in the 
dusty ERG population. Given the lower redshift of PDFJ011423, this source 
offers the best opportunity to study the interplay between the two 
phenomena in the dusty environments of these galaxies.

The ratio of the radio power to the dust-corrected optical luminosity of 
PDFJ011423, $P_{1.4}/L_R \sim 1$, confirms that PDFJ011423 is radio 
quiet. It seems likely that this source is linked to the radio-quiet 
counterparts of the red quasar population discovered by \citet{Web95} and 
also to the ERGs recently discovered among the population responsible for
the hard X-ray background \citep{Has01}. Future work will show if the 
intense star formation activity present in PDFJ011423 is also a common 
feature to the above mentioned obscured AGN populations. 

All these different studies suggest the existence of an important population 
of dusty ERGs, powered by heavily obscured starbursts and/or AGNs, which 
is now starting to be observed. Given their extreme nature, 
the study of the dusty ERGs will hold fundamental clues to the 
understanding of galaxy evolution, by revealing valuable information
on the hidden star formation and AGN activity in the universe.

\acknowledgments

We are grateful to A. Verma for adapting and running her SED fitting 
algorithm for this source and for useful discussions. We also thank A. 
Efstathiou and M. Rowan-Robinson for their advice. The award of the 
European Southern Observatory Director's discretionary time is 
gratefully acknowledged. JA gratefully acknowledges a scholarship from 
the Science and Technology Foundation (FCT, Portugal) and the assistance 
from the University of London Valerie Myerscough Prize. BC and LC 
acknowledge the support of the Australian Research Council.

\clearpage

\begin{deluxetable}{cccc}
\tablewidth{0pc}
\tablecaption{Line measurements for PDFJ011423 \label{tab:lines}}
\tablehead{
\colhead{Line} & \colhead{$\lambda_{\rm obs}$} &
\colhead{F$_{\rm obs}$} & \colhead{$W_{\lambda,\rm rest}$}\\
\colhead{} & \colhead{(\AA)} &
\colhead{(10$^{-16}$ erg s$^{-1}$ cm$^{-2}$)} & \colhead{(\AA)} }
\startdata
[OII]$\lambda \lambda$3726,3729 & $6159 \pm\ 1$ & $8.0 \pm\ 0.3$ 
& $90 \pm\ 6$  \\

H$\beta$                        & $8033 \pm\ 4$ & $1.0 \pm\ 0.2$ 
& $10 \pm\ 4$ \\

[OIII]$\lambda$4959             & $8195 \pm\ 4$ & $2.0 \pm\ 0.8$ 
& $19 \pm\ 11$ \\

[OIII]$\lambda$5007             & $8274 \pm\ 4$ & $3.7 \pm\ 0.7$ 
& $32 \pm\ 10$ \\

H$\alpha$+[NII]\tablenotemark{a} & $10858 \pm\ 11$ & $28.7 \pm\ 2.5$ 
& $217 \pm\ 47$  \\

\enddata
\tablenotetext{a}{Measurements from a single Gaussian fit to the observed 
blend.}

\end{deluxetable}

\clearpage

\begin{deluxetable}{lcccr}
\tablecolumns{5}
\tablewidth{0pc}
\tablecaption{Comparison  
between the observed properties of dusty ERGs\label{tab:ld}}
\tablehead{
\colhead{}& \colhead{PDFJ011423} & \colhead{HR10\tablenotemark{1}} 
& \colhead{ISOJ1324-2016\tablenotemark{2}} 
& \colhead{EROJ164023}\tablenotemark{3} }
\startdata
$z$   & 0.65 & 1.44 & 1.50 & 1.05 \\ 
$K$ mag & 15.3 & 18.4 & 17.5 & 17.6 \\ 
$R-K$  & 5.8 & \nodata & $>5.2$ & 5.9 \\ 
$I-K$  & \nodata & 7.8 & 4.9 & 4.5 \\ 
$J-K$  & 3.1 & 2.6 & \nodata & 2.1 \\
L$_{\rm{H}\alpha}$\,(W) & $2.0-2.6\times 10^{35}$ 
& $2.6 \times 10^{35}$ & \nodata &  
$0.7 \times 10^{35}$ \\ 
L$_{\rm{H}\alpha, corr}$\,(W)\tablenotemark{\dagger} & $0.9-2.6\times 10^{37}$ 
&  \nodata & \nodata & \nodata \\ 
L$_{10\,\mu\rm{m}-2\,\rm{cm}}$ (L$_{\sun}$)  &  $6.5\times 10^{12}$ & 
$4.0\times 10^{12}$ & \nodata & 
$\lesssim 2\times 10^{12}$ \\ 
SFR (M$>5$M$_\sun$, M$_\sun$\,yr$^{-1}$)  & $\sim 470$ & 
$200-400$ & \nodata & $\sim 6-400$\\ 
M$_{dust}$ (M$_{\sun}$)  & $8-30 \times 10^{8}$ & 
$\sim 4\times 10^{8}$ & \nodata & \nodata  \\ 
$A_V$ (mag) & $5-6$ & $2-4.5$ & $4-7$ & $\sim 5$ \\ 
\enddata
\tablenotetext{\dagger}{H$\alpha$ luminosity corrected for dust extinction.}
\tablecomments{All the values are given for the cosmology adopted in this 
paper.}
\tablerefs{(1) Dey {\it et al.} 1999; (2) Pierre {\it et al.} 2001; 
(3) Smith {\it et al.} 2001} 
\end{deluxetable}

\clearpage

\end{document}